\documentclass[journal=langd5,manuscript=article]{achemso}

\usepackage[version=3]{mhchem} % Formula subscripts using \ce{}
\usepackage{graphicx,color}% Include figure files
\usepackage{dcolumn}% Align table columns on decimal point
\newcolumntype{.}{D{.}{.}{-1}}
\usepackage{bm}% bold math
\usepackage[latin1]{inputenc}
\usepackage{color}

\newcommand{\lmnp}{LiMnPO$_{4}$}
\newcommand{\lfp}{LiFePO$_{4}$}

\newcommand{\LiPO}{Li\(_3\)PO\(_4\)}
\newcommand{\MnOHPO}{Mn\(_2\)(OH)PO\(_4\)}

\author{Christoph Neef}
\email{christoph.neef@kip.uni-heidelberg.de}
\author{Carsten J\"ahne}
\affiliation[University of Heidelberg]
{Kirchhoff Institute for Physics, University of Heidelberg, D-69120 Heidelberg, Germany}
\author{Hans-Peter Meyer}
\affiliation[University of Heidelberg]
{Institut f{\"u}r Geowissenschaften, University of Heidelberg, D-69120 Heidelberg, Germany}
\author{R{\"u}diger Klingeler}
\email{ruediger.klingeler@kip.uni-heidelberg.de}
\affiliation[University of Heidelberg]
{Kirchhoff Institute for Physics, University of Heidelberg, D-69120 Heidelberg, Germany}

\title{Morphology and agglomeration control of LiMnPO$_4$ micro- and nanocrystals}

\begin{document}

\begin{abstract}
Microwave-assisted hydrothermal synthesis was used to grow LiMnPO$_4$ micro- and nanocrystals from acetate precursors.
By appropriate adjustment of the precursor concentration and the pH-value of the reactant, the
product composition and purity along with the crystal size can be manipulated, resulting in
particle-dimensions from around 10 $\mu$m down to a few 100 nm. Prisms and plates with hexagonal
basal face as well as cuboid and rod-like particles were produced. The effects on the crystal morphology as well as on
the materials texture and agglomeration tendency are discussed and a comprehensive agglomeration phase
diagram is constructed.
\end{abstract}

\section{Introduction}

The properties of crystalline nanomaterials are intimately connected with their size, shape, and microstructure so that
downsizing and tailoring the morphology of materials provides control about, e.g., their electronic, magnetic, and
optical properties.~\cite{Vollath,Kelly} In addition to the properties of  primary particles, their tendency to form
aggregates is highly relevant for tailoring desired materials optimized regarding the basic physico-chemical parameters
as well as, e.g., their overall toxicity, reactivity, and mechanical stability. Therefore, understanding and
controlling accretion of nanocrystals and their oriented or random agglomeration to larger, possibly textured entities
is a precondition for adjusting a particular property for the desired application.~\cite{Penn} In addition, research
activities do not only focus on improving materials for particular applications but include the issue of cost and
energy efficient production routes for environmentally benign materials.~\cite{Guo2008,Zakharova2012}.

A particularly relevant application of nanostructured materials is their usage as electrode materials in
electrochemical energy storage devices, where, e.g., Li-diffusion, electronic conductivity, structural stability have
to be optimized~(see e.g.~\cite{Bruce,Armand2002,Popa}). In particular, reduction of the ionic and the electronic
diffusion lengths and increase of the surface to volume ratio often yields significant improvements in the
electrochemical activity. This is particularly illustrated by the olivine phosphates Li\textit{M}PO$_4$ with
\textit{M}=(Mn, Fe, Co, Ni)~\cite{Padhi1997,Yamada2001,Amine2000,Okada2001} which can be applied in lithium-ion
batteries nano-sized only. The olivine \lmnp\ is a promising cathode material owing to the Mn$^{2+}$/Mn$^{3+}$ redox
potential at 4.1 V vs. Li and the associated flat discharge curve~\cite{Kim2010}. Stable reversible capacities of up to
145 mAh/g have already been obtained in carbon-coated nanostructured material consisting of $\sim 30$\,nm thick
platelets which extend several 100\,nm in the base plane~\cite{HpnW} while the performance of rather spherical
particles decreases significantly for particle sizes of $\sim$200\,nm.~\cite{Drezen} In addition to the bare size
reduction of the particles, the particular morphology is relevant as well, since the ionic diffusivity was predicted
not only small but also strongly anisotropic in bulk materials favoring the crystallographic
$b$-direction~\cite{Morgan}. Experimental evidence for this anisotropy is indeed deduced from recent studies on
nanostructured materials~\cite{HpnW,GsaJ} and from single crystals~\cite{Wizent}. These results imply that
non-spherical crystals with shortest dimensions along the crystallographic direction(s) of high ionic conductivity are
optimal for actual applications in lithium-ion batteries. Furthermore, the orientation and the contact of the particles
with respect to each other, to the electrolyte, and to the current collector have notable influence on the
electrochemical performance of a material.~\cite{Li2006} Therefore, the particles surface properties must not only be
adjusted to meet the demands of conductivity and chemical stability but also of the particle distribution and
agglomeration, which are governed by surface charging effects. In addition to their applicability for energy storage,
the olivine phosphates exhibit antiferromagnetic spin order and a large magnetoelectric effect in the magnetically
ordered phase which is of fundamental research interest but may yield to future applications, too.~\cite{Toft} This is
highlighted by the recent discovery of unusual ferrotoroidic domains in LiMPO4 with M = Co or Ni which is discussed in
terms of future data storage devices.~\cite{Fiebig}

In this work, we investigate the tailored synthesis of dispersed and agglomerated \lmnp\ with dimensions between
$\sim$100\,nm and several $\mu$m aiming at fundamental information which enables tailoring nano- and microscaled
material. We demonstrate that control of the size and shape of \lmnp\ crystals as well as of the particles' tendency to
(oriented) agglomeration is possible by applying a fast and low-energy synthesis route, starting from environmentally
harmless and non-hazardous precursors. The microwave-assisted hydrothermal method applied here allows controlled
synthesis at relatively low temperatures with a non-toxic solvent. Because of their good solubility in water and
absence of any additional environmental impact, acetates were selectively chosen as cation-precursors. The phosphate
source (di-ammonium hydrogen phosphate) is commonly used as fertilizer in agriculture. Besides ammonia, no further
organic additives were used in the educt, thus rendering the chosen synthesis route elementary and reproducible.

\section{Experimental}

Applying a hydrothermal synthesis technique, \lmnp\ was synthesized using lithium acetate dihydrate, manganese(II)
acetate tetrahydrate, and di-ammonium hydrogen phosphate (Aldrich, $\geq$99\%) solved in deionized water in an
off-stoichiometric ratio of 3:1:1 with respect to the ratio of the Li-, Mn-, and PO$_4$-ions. The concentration of the
Mn-ions in the solution was adjusted to a value between 0.03 and 0.13 mol/l. The pH-value of the reactant was measured
with a SevenGo SG2 system (Mettler Toledo) and, starting from a value of 5.5-5.8, adjusted to a value of up to 11.5 by adding diluted ammonia (AppliChem).

10~ml of the solution was put into a 30~ml glass reaction vessel equipped with a magnetic stir bar which
was transferred into a Monowave 300 microwave reactor (Anton-Paar) with 2.46~GHz radiation.~\cite{Jaehne} Within a
ramping time of 10~min the vessel was heated to a fixed temperature between 160~{°C} and 220~{°C} and held at this
temperature for 20~min. Subsequent cooling was carried out by the use of compressed air. Note, that high precursor concentrations and superalkaline pH-values could not be achieved
simultaneously due to pressure limitations of the device since the vapor pressure of the ammonium adds to that of the other educts.
After the synthesis procedure, the precipitate was recovered from the vessel, washed several times with deionized water and ethanol and dried at 90~°C over night.

The energy efficiency of the applied microwave system can be roughly estimated by considering the radiation power for
one of the synthesis procedures. The total energy deposit upon heating up the reaction chamber and the educts amounts
to about 28~kJ. On the other hand, heating 10~ml of water and the glass reaction vessel (54~g) from 20~°C to 220~°C
consumes $\approx$8.4~kJ (+0.3~kJ vaporization heat) and $\approx$7.6~kJ, respectively. Neglecting any effects due to
the chemical reaction hence implies that the efficiency of this laboratory system is roughly about 50\%. Note, that
this value may easily be increased when up-scaling and optimizing the process so that the method offers very high
efficiency compared to any ceramic or conventional hydrothermal procedures.

X-Ray powder diffraction (XRD) was performed in Bragg-Brentano geometry (Siemens D500) applying Cu-K$_{\alpha1}$ and
Cu-K$_{\alpha2}$ radiation ($\lambda_{1/2} =  1.54056 / 1.54433$~\AA). The measurements were taken in the 2 $\theta$
range from 10° to 70° with a step size of 0.02° and integration time of 1~s (short) or 10~s (long) per step. Prior to
the measurements, the powders were dispersed in isopropyl alcohol and dripped on a glass sample carrier. Structural
phase analysis was conducted with the FullProf Suite 2.0 program \cite{fullprof}. The morphology of the material was
studied by means of a ZEISS Leo 1530 scanning electron microscope with an acceleration voltage of 9~kV. Preliminarily,
the samples were coated with a thin gold layer of about 10~nm with a Balzers Union SCD 004 sputtering device under
Argon atmosphere.

\section{Results and discussion}

\subsection{Synthesis parameters}

Solutions of the individual precursors are pellucid and show a pH-value of 8.1 ((CH$_3$COO)Li), 7.5 ((CH$_3$COO)$_2$Mn)
and 8.1 ((NH$_4$)$_2$HPO$_4$), which is a result of the soft basic properties of CH$_3$COO$^{-}$ and HPO$_4^{2-}$ ions.
A white precipitate is formed upon mixing of dissolved (NH$_4$)$_2$HPO$_4$ and (CH$_3$COO)$_2$Mn, which is accompanied
by a drop of pH-value to 5.5. The addition of (CH$_3$COO)Li has little effect. No color change can be observed upon
alkalizing the solution with ammonia up to a pH of around 10.7. Above this value the color changes to beige and
eventually brown, possibly because of the oxidation of Mn$^{2+}$ to manganite or manganese dioxide with air oxygen.

In order to investigate the white precipitate, solutions with pH of 5.5 (starting solution), 9 and 11 were dried and
the products were measured by means of XRD (see the supporting information). All three samples exhibit the diffraction
pattern typical for NH$_4$MnPO$_4\cdot$H$_2$O.~\cite{Carling1995} However, the XRD patterns of the samples prepared at pH
5.5 and 11 show a high background, which points to the presence of amorphous phases. The sample prepared at pH 9
exhibits sharp peaks and a low background which is characteristic for a good crystallinity. All samples show a high
degree of particle orientation in the (010) plane which is most pronounced in the sample prepared at pH 9. This
predominant alignment can be explained by the plate-like morphology of the obtained crystallites (for electron
micrographs see the supporting information).

\begin{figure}
\includegraphics[width=0.95\columnwidth,clip]{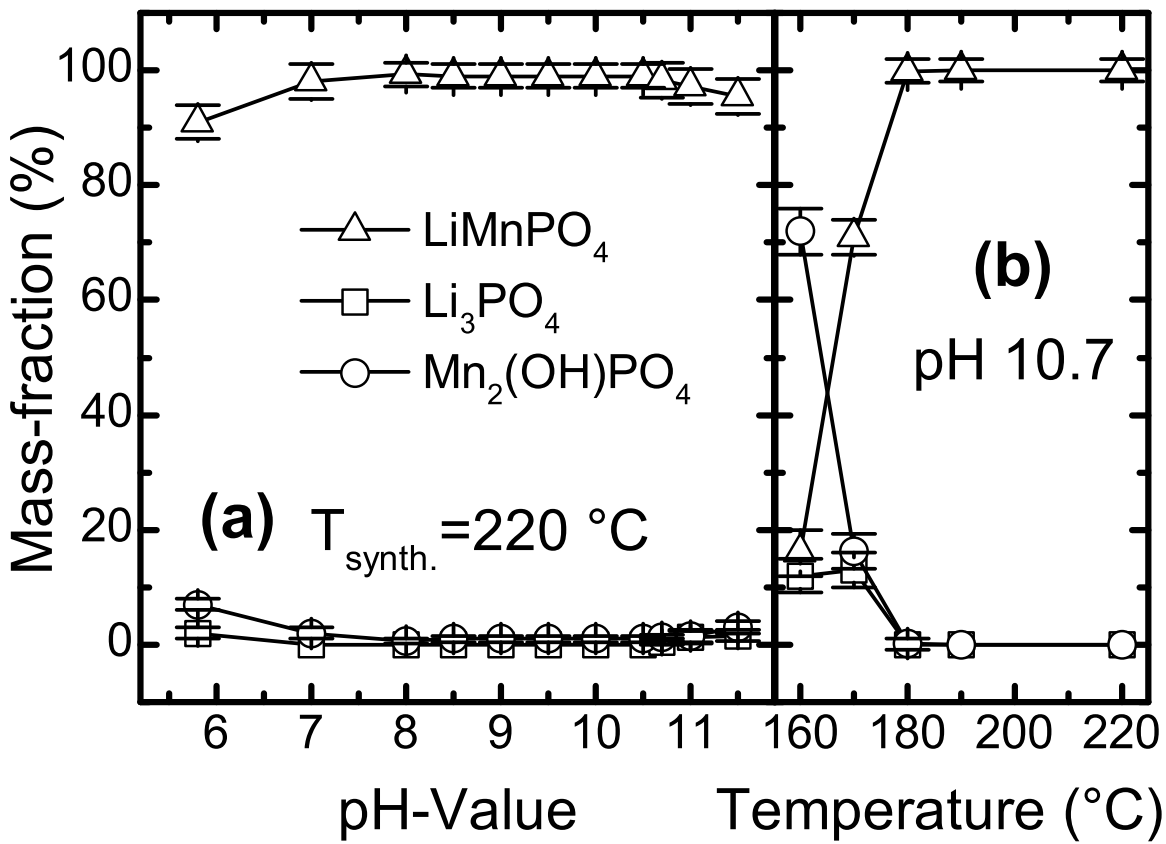}
\caption{(a) Weight fraction of LiMnPO$_4$, Li$_3$PO$_4$ and Mn$_2$(OH)PO$_4$ in dependence of (a) the educt pH-value at
fixed temperature and (b) the synthesis temperature at fixed pH-value as deduced from Rietveld fits to the XRD data.}
\label{parameterraum}
\end{figure}

The subsequent hydrothermal syntheses as described above yield white to beige colored powders. The pH of the reaction
products as recovered from the microwave reactor undermatched the pH of the solutions before reaction by an average
value of 1. All washed and dried materials were studied by powder XRD and characteristic patterns of the olivine Pnma
structure realized in \lmnp\ were found (see the supporting information). A rough inspection of the data shows that
several samples exhibit \LiPO\ and \MnOHPO\ phases in addition to the main phase. Hence, a continuative phase analysis
was made for each sample by executing a basic multi-component phase refinement of the XRD patterns by fitting the mass
fractions of the three compounds LiMnPO$_4$ (ICSD \#97763 \cite{IotG}), \LiPO\ (ICSD \#25816 \cite{DkvZ}), and \MnOHPO\
(ICSD \#16928 \cite{CsoW}) with pseudo-Voigt profile functions. The results of this analysis for materials synthesized
with a fixed precursor concentration of 0.03~mol/l with respect to the Mn$^{2+}$-ions in the starting solution are
shown in Fig.~\ref{parameterraum} for various pH-values of the reactant and synthesis temperatures.

%%%%%%%%%%%%%%%%%%%%%%%%%%%%%%%%%%%%%%FIG%%%%%%%%%%%%%%%%%%%%%%%%%%%%%%%%%%%%%%%%%%%%%%%%%%%%%
%%%%%%%%%%%%%%%%%%%%%%%%%%%%%%%%%%%%%%%%%%%%%%%%%%%%%%%%%%%%%%%%%%%%%%%%%%%%%%%%%%%%%%%%%%%%%%

The data in Fig.~\ref{parameterraum} (a) show that, at the synthesis temperature of 220~°C, the production of phase
pure LiMnPO$_4$ without formation of side products is possible in a wide pH-range between 7 and 10.7. Below and above
this regime the materials exhibit \LiPO\ and \MnOHPO\ impurity phases in the \%-range which can be detected by the XRD
experiment. Note, that most of the overview XRD patterns were measured with a short integration time so that the error
bars are in the range of a few mass-percent. The XRD patterns discussed in the next sections were measured more
accurately.

The regime of stable LiMnPO$_4$ production coincides with the pH-range favoring crystalline NH$_4$MnPO$_4\cdot$H$_2$O
without amorphous side products. This compound is known to be a precipitate precursor at the synthesis of LiMnPO$_4$
due to its structural similarities.~\cite{Bramnik2007,Liu2012} We thus assume that NH$_4$MnPO$_4\cdot$H$_2$O is a
necessary precursor for the synthesis of LiMnPO$_4$ under these conditions. The residual amorphous phases present in
the educt are likely to be reacted to the impurity phases \LiPO\ and \MnOHPO . Note, that calculations in
Ref.~\cite{Delacourt} suggest that a production of Li$_3$PO$_4$ is more likely at superalkaline conditions which favor
the phosphate form PO$_4^{3-}$ in aqueous solutions. This conclusion is supported by the results at hand.

In Fig.~\ref{parameterraum} (b), the effect of synthesis temperature variation is illustrated for fixed pH = 10.7. The
data show that, under these conditions, phase pure synthesis of LiMnPO$_4$ occurs at synthesis temperatures exceeding
180~°C. In general the results in Fig.~\ref{parameterraum} demonstrate that with the chosen precursors the experiment
offers a wide range of parameters where \lmnp\ can be successfully synthesized while their variation may be used to
control the morphology and the size of the produced particles. In the following, the synthesis temperature was chosen
to be 220~°C in all cases.

\subsection{Particle size and morphology}

%%%%%%%%%%%%%%%%%%%%%%%%%%%%%%%%%%%%%%FIG%%%%%%%%%%%%%%%%%%%%%%%%%%%%%%%%%%%%%%%%%%%%%%%%%%%%%
\begin{figure}
\includegraphics[width=0.95\columnwidth,clip]{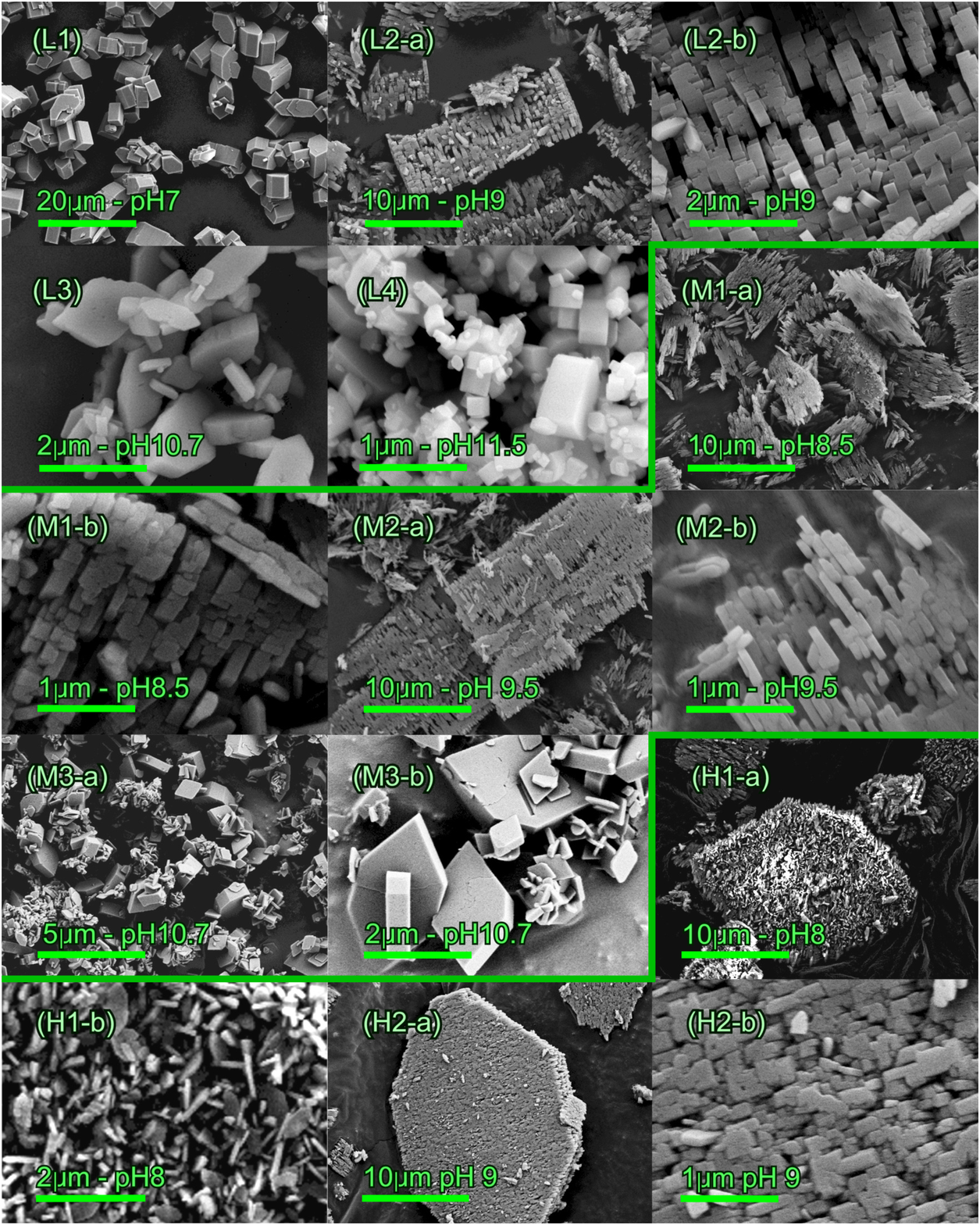}
\caption{Electron micrographs of LiMnPO$_4$ samples, synthesized with three different precursor concentrations and various
pH-values (see the text). \textbf{L}: 0.03 mol/l Mn$^{2+}$-ions; \textbf{M}: 0.075 mol/l Mn$^{2+}$-ions; \textbf{H}:
0.13 mol/l Mn$^{2+}$-ions.} \label{SEM}
\end{figure}
%%%%%%%%%%%%%%%%%%%%%%%%%%%%%%%%%%%%%%%%%%%%%%%%%%%%%%%%%%%%%%%%%%%%%%%%%%%%%%%%%%%%%%%%%%%%%%

Fig.~\ref{SEM} gives an exemplary overview on the LiMnPO$_4$ morphologies accessible by the applied synthesis route.
The presented electron micrographs are grouped according to the three different precursor concentrations (\textbf{L}:
0.03~mol/l Mn$^{2+}$-ions; \textbf{M}: 0.075~mol/l Mn$^{2+}$-ions; \textbf{H}: 0.13~mol/l Mn$^{2+}$-ions) which were
used. Within each group, the images are arranged with respect to the pH-value of the reactant before the synthesis.

The largest LiMnPO$_4$ crystals are obtained at a low precursor concentration and neutral to alkalescent pH-values.
Sample L1 consists of hexagonal prisms with a base-length of about 10~$\mu$m and a height of 3 to 5~$\mu$m. As
demonstrated by the following images, the hexagonal basal face is a typical element of hydrothermally synthesized
LiMnPO$_4$ particles (cf. also Ref.~\cite{Chen}). Increasing the pH-value to 9 (sample L2) affects both the crystal
size and the agglomerative behaviour of the particles. The material features rectangular aggregates with an ordered
texture and consisting of sub-$\mu$m-sized cuboidal particles. This strong tendency to aggregation disappears when the
reactant is heated at even more alkaline conditions. Sample L3 produced at pH 10.7 contains arbitrarily arranged
platelets, again with a hexagonal however rounded basal face. The longest edge amounts 1 -- 2~$\mu$m and the platelets
heights are sub-$\mu$m. The smallest particles obtained at this precursor concentration were produced at pH of 11.5
which is the highest pH-value applied. While the morphology of the resulting particles is similar to sample L1, the
crystals are sub-$\mu$m to nano-sized in any direction. Note, however, that at this strong alkaline conditions no phase
pure LiMnPO$_4$ synthesis was possible (see Fig.~\ref{parameterraum}).

The above described tendency to agglomeration into ordered textures intensifies when increasing the precursor
concentration. At pH-value of 8.5 (sample M1) the process yields plates with dimensions of about 10~$\mu$m consisting
of cuboidal particles with a typical size of some 100~nm. A similar hierarchically assembled structure is realized in
sample M2, synthesized at pH 9.5. When increasing the pH-value, the size of the agglomerated structures extends.
Concomitantly, growth of the individual particles seems to favour a particular direction which yields rod-like crystals
with a typical length of about 0.5~$\mu$m being the elements of the agglomerates (see images M2-a and M2-b). Further
raise of alkalinity to pH = 10.7 (sample M3) stops the strong aggregative tendency. Sections M3-a and M3-b of
Fig.~\ref{SEM} display well-developed hexagonal platelets, however with a very broad size distribution with in-plane
particle dimensions from about 2~$\mu$m down to some 100~nm.

At the highest precursor concentration of 0.13~mol/l Mn$^{2+}$-ions, textured agglomerates are observed in the whole
accessible pH-range. All materials exhibit well-ordered agglomerates with hexagonal to octagonal shape and sizes larger
than 10~$\mu$m. The electron micrographs of sample H1, synthesized at an educt pH-value of 8, indicates that the
aggregates consist of non- or low-ordered sub-$\mu$m particles. Note, that the structure analysis described below
reveals that the crystallites exhibit an ordered texture, thus rendering the apparent disorder surface related. This is
more clearly visible in the micrographs H2-a and b of the material produced at pH 9. Here, the ordered texture of the
agglomerates surface formed by cuboidal particles with a typical size of some 100~nm is well visible. Interestingly,
the flake-like aggregates can be multilayered, thereby forming 'agglomerates of agglomerates'. Note, that neither the
morphology of the primary particles nor of the aggregates changes when the educt pH is increased 10.4 (not shown).

%%%%%%%%%%%%%%%%%%%%%%%%%%%%%%%%%%%%%%FIG%%%%%%%%%%%%%%%%%%%%%%%%%%%%%%%%%%%%%%%%%%%%%%%%%%%%%
\begin{figure}
\includegraphics[width=0.95\columnwidth,clip]{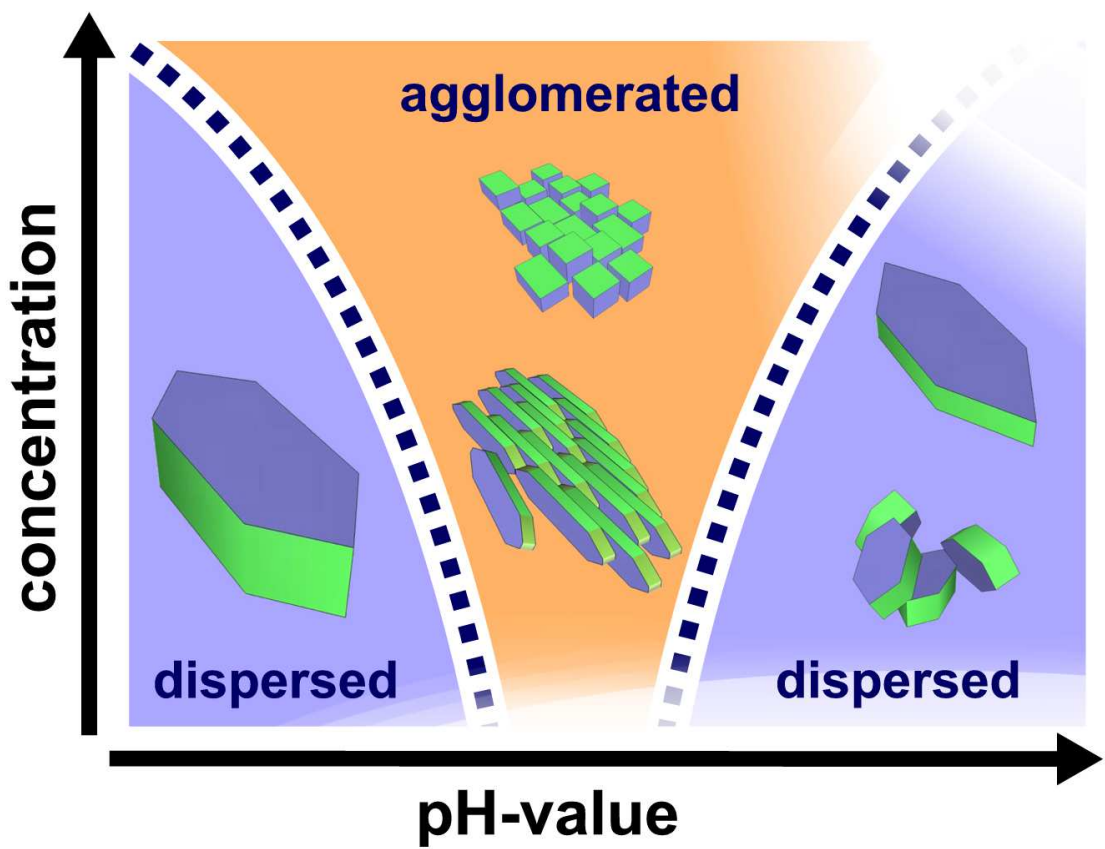}
\caption{Schematic sketch of the pH-value and precursor concentration dependence of particle-size and -agglomeration. The regime of stable dispersion is stained blue, the regime of ordered agglomeration is stained orange.}
\label{sizepattern}
\end{figure}
%%%%%%%%%%%%%%%%%%%%%%%%%%%%%%%%%%%%%%%%%%%%%%%%%%%%%%%%%%%%%%%%%%%%%%%%%%%%%%%%%%%%%%%%%%%%%%

Summarizing, variation of the precursor concentration and the pH-value of the reactant strongly affects the size and
shape of the individual crystallites and of the agglomerates. Note, that the particles morphologies obtained after
hydrothermal conditions are not related to the morphology of the crystalline NH$_4$MnPO$_4\cdot$H$_2$O precursor which
was recovered from the starting solution by conventional drying.

The results are illustrated schematically in Fig.~\ref{sizepattern}. In general, the particle size decreases upon
raising alkalinity of the reactant and increasing the precursor concentration. The latter yields an increase of the
number of condensation nuclei which explains the smaller particle sizes. This behaviour is however modified by
simultaneous variation of the particles tendency to agglomerate into larger entities, which also depends both on the
pH-values of the solutions and the precursor concentrations. Agglomeration is usually favoured by attractive van der
Waals interactions and prohibited by the electrostatic surface potential. In Fig.~\ref{sizepattern}, the region of
textured agglomerates is indicated by orange colour. In contrast, blue colour signals the parameter space where
individual particle formation is favoured.

At highest pH, the particles are medium-sized and show low tendency to clot to large structures. Presumably, the strong
alkalinity of the solution causes a finite negative surface electrostatic potential preventing contacts of the
relatively low concentrated seeds and particles. Straightforwardly, both increasing the precursor concentration and
decreasing pH stabilizes the coagulation processes which is visible in Fig.~\ref{sizepattern}. In case of neutral
surfaces which in the case at hand seems to be realized between pH 8 and 9.5, agglomeration should occur regardless of
the educt concentration (see, e.g., Ref.~\cite{Rector,Li2007}) which is indeed observed in the experiment. Further acidising
is supposed to generate positively surface charged and hence well dispersed particles. In addition to appropriate
charge stabilization in water a narrow primary particle size distribution is necessary for the formation of ordered
agglomerates, too~\cite{Mileni2007}. Indeed, the supporting effect of adapted primary particle size and shape is
evident if the strong agglomeration regime in Fig.~\ref{SEM} is compared to the region of dispersed particles.

Note, that very recently the isoelectric point in \lfp\ was found at pH = 4.3 \cite{Li2012}. Firstly, the apparent
deviation from pH between 8 and 9.5 which is observed in our experiments for LiMnPO$_4$ needs to be qualified by a
small uncertainty of the pH determination during particle formation. After the syntheses, the pH-values were slightly
lower (by a maximum of 1) as compared to the educts which is due to the release of NH$_4^{+}$-ions from the
NH$_4$MnPO$_4\cdot$H$_2$O precipitate. Acidic conditions however were never reached in our experiments so that there is
a clear contrast to \lfp. In iron phosphate, acidic groups resulting from unsaturated metallic cations are supposed to
dominate the surface properties in aqueous solution leading to a negative surface charge for pH-values even below
3~\cite{Lee2010}. This coincides with the isoelectric point at acidic pH-values. In contrast, our data presented above
show that the dominant surface groups of LiMnPO$_4$ are less acidic. We suggest that a lower attraction of
OH$^{-}$-ions in LiMnPO$_4$ as compared to LiFePO$_4$ causes the shift of the isoelectric point to the basic regime
which is consistent with lower acidity of aqueous Mn$^{2+}$-ion as compared to Fe$^{2+}$ (pK$_a$(Mn$^{2+}$) = 10.59,
pK$_a$(Fe$^{2+}$) = 9.5).

\subsection{Structure analysis and preferred orientation effects}

%%%%%%%%%%%%%%%%%%%%%%%%%%%%%%%%%%%%%%FIG%%%%%%%%%%%%%%%%%%%%%%%%%%%%%%%%%%%%%%%%%%%%%%%%%%%%%
\begin{figure}
\includegraphics[width=0.95\columnwidth,clip]{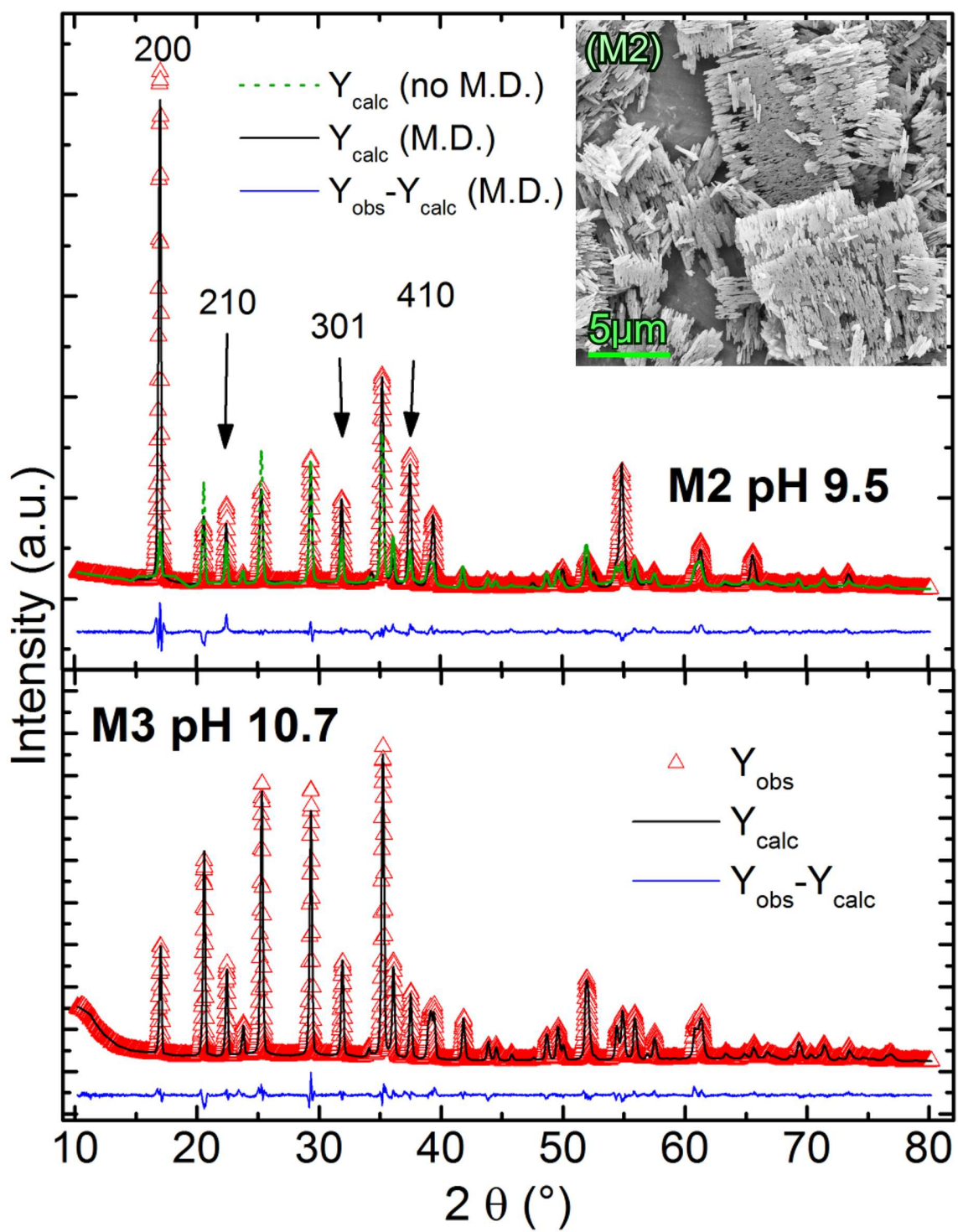}
\caption{Measured and calculated XRD patterns for sample M2 and M3. The labeled peaks for M2 are enlarged as compared
to the calculated pattern (green line). This effect is considered by means of the March-Dollase model (black line).
Inset: SEM picture of sample carrier after wet preparation process.} \label{xrd}
\end{figure}
%%%%%%%%%%%%%%%%%%%%%%%%%%%%%%%%%%%%%%%%%%%%%%%%%%%%%%%%%%%%%%%%%%%%%%%%%%%%%%%%%%%%%%%%%%%%%%

A more detailed structural analysis was conducted for two samples (M2, M3) synthesized at a moderate precursor
concentration of 0.075 mol/l and at pH-values of 9.5 and 10.7, respectively, representing the two different cases of
highly agglomerated and dispersed particles. Fig.~\ref{xrd} shows the measured XRD patterns of both samples along with
the calculated pattern (with pseudo-Voigt profile functions) according to Ref.~\cite{IotG}, including the March-Dollase
Model, and the resulting residual. Both patterns exhibit the diffraction peaks characteristic for the orthorhombic Pnma
structure, however huge differences in the relative peak intensities between both samples are visible. While sample M3
shows the pattern of arbitrarily oriented particles expected for dispersed particles, especially the (200), (210),
(301), (410), and (610) peaks at $2\theta$ = 17.0°, 22.4°, 31.8°, 37.5°, and 54.9° are enlarged in sample M2 as
compared to Ref.~\cite{IotG}. This effect is well known for oriented powders and the aberration can be modeled by
applying the March-Dollase Model \cite{CoiD} for a preferred powder orientation, with the March-Dollase function:

\begin{equation}
     P(r,\alpha) = (r^2 cos^2(\alpha) + r^{-1} sin^2(\alpha))^{-3/2}.
\label{MarchDollase}
\end{equation}

$P(r,\alpha)$ describes the volume fraction of the material oriented at an angle $\alpha$ with respect to the
$\hat{n}$-axis, which is normal to a specific crystallographic plane (hkl), and a parameter $r$. Refinement of $r$ in
the accordingly modified pattern with the (200), (210), and (301) reflection yields the coefficients listed in
table~\ref{PrefMD}. Considering the (hkl)-planes (410) and (610) does not further improve the agreement between model
and measured data. Note, that only the orientation, not the structure was refined.

%%%%%%%%%%%%%%%%%%%%%%%%%%%%%%%%%%%%%%%%%%%%%%%%%%%%%%%%%%%%%%%%%%%%%%%%%%%%%%%%%%%%%%%%%%%%%%
\begin{table}[htbp]
             \caption{Calculated March-Dollase multi-axial model parameters for sample M2.}
            \begin{tabular}{lcc}
            \hline
            \textbf{Pref. (hkl)} \hspace{3mm} &  \textbf{$\hat{n}$-fraction} \hspace{2mm} & \textbf{$r$} \hspace{2mm} \\
            (200) & 0.936 & 0.503 \\
            (210) & 0.054 & 0.441 \\
            (301) & 0.009 & 0.308 \\
            \hline
        \end{tabular}
     \label{PrefMD}
\end{table}
%%%%%%%%%%%%%%%%%%%%%%%%%%%%%%%%%%%%%%%%%%%%%%%%%%%%%%%%%%%%%%%%%%%%%%%%%%%%%%%%%%%%%%%%%%%%%%

\begin{figure}
\includegraphics[width=0.95\columnwidth,clip]{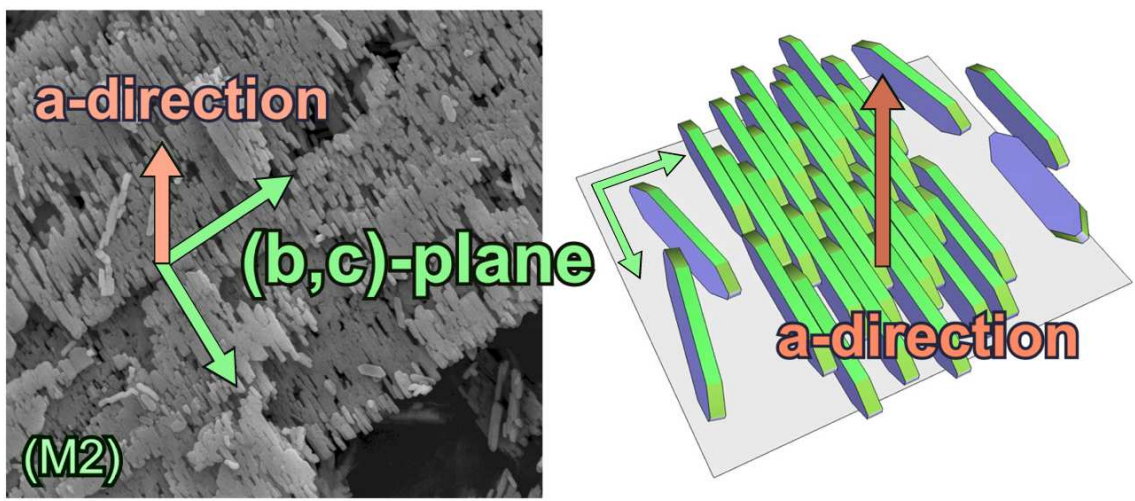}
\caption{Relation between agglomeration behaviour of primary particles and crystallographic directions. The ($b$,$c$)-surface is green coloured, all perpendicular surfaces are blue coloured.}\label{SEM3D}
\end{figure}

The obtained parameters show that the primary crystals are mainly aligned in the (100) plane. This corresponds to the
orientation of the crystallographic \textit{bc}-plane perpendicular to the plane generated by the X-ray beam and thus
parallel to the sample holder. This outcome might be caused by the interplay of two effects, i.e. (1) an anisotropic
tendency to agglomeration of the primary nanocrystals, and (2) the arrangement of the plate-like aggregates of sample
M2 parallel to the sample carrier (see the inset of Fig.~\ref{xrd}). The latter might result from the wet preparation
process. In this way, the alignment of the primary particles leads to the fixing of one preferred axis (see
Fig.~\ref{SEM3D}). The specific orientation of the crystallographic $b$- and $c$-directions can not be deduced from
these data. A comparison with electron diffraction data from Ref.~\cite{Chen} however suggests that the direction of
longest particle dimension corresponds to the $c$-axis.

The results imply anisotropic agglomeration tendencies between the \textit{bc}-surfaces (no or very low tendency to
agglomeration) and the \textit{ab}- and \textit{ac}-surfaces, respectively (strong agglomeration), and hence indicate
different tendencies to charge accretion and thus different repulsive forces. To be specific, the results imply higher
charging of the \textit{bc}-surface even in the regime of high agglomeration as compared to the other surfaces.

From a structural point of view, the \textit{bc}-planes are formed by layers of corner-sharing MnO$_6$ octahedra
separated by layers containing the PO$_4$ tetrahedra and Li-channels. Whereas, in the \textit{ab}- and
\textit{ac}-planes only layers containing different polyhedra can be found (for illustration see for example
\cite{Bramnik2007}). The higher charging of the \textit{bc}-surface might hence result from the presence of just one
species at the surface and thus the accretion of only ions with the same sign of charge, e.g., Mn-hydroxocomplexes
formed by unsaturated ions in the Mn-oxide layer (acidic behavior) or the formation of hydrogen phosphate with
PO$_4^{3-}$ groups (basic behavior).

The observed powder XRD orientation effect can be seen to some extend in the patterns of all materials which exhibit
the tendency to form ordered agglomerates. In this respect, the analysis of the XRD data confirms the general trends in
agglomeration summarized in Fig.~\ref{sizepattern}.

Diverse surface functionalities were artificially applied by Xia et al. \cite{Xia2009} to Ag-nanocubes dispersed in
water. Their observation of self-assembled structures confirms a significant influence of the surface properties on the
growth, the texture, and the final size of the agglomerates. Similarly, as outlined above we propose that the formation
of diversely shaped and partly oriented agglomerates found in LiMnPO$_4$ nanostructures is governed by the particular
surface properties. Note, that a different mechanism is proposed for LiFePO$_4$ microstructures which have been
recently synthesized via a hydrothermal route.~\cite{Su2012} While similar to our findings self-assembled and ordered
agglomerates of LiFePO$_4$ nanocrystallites are observed, Su \textit{et al.} emphasize the relevance of the reaction
time and of the template properties of the precursors for the particular morphology of the aggregates while the
specific effect of surface charge is not taken into consideration.

%%%%%%%%%%%%%%%%%%%%%%%%%%%%%%%%%%%%%%FIG%%%%%%%%%%%%%%%%%%%%%%%%%%%%%%%%%%%%%%%%%%%%%%%%%%%%%
%%%%%%%%%%%%%%%%%%%%%%%%%%%%%%%%%%%%%%%%%%%%%%%%%%%%%%%%%%%%%%%%%%%%%%%%%%%%%%%%%%%%%%%%%%%%%%

\subsection{Summary}

We have demonstrated the influence of the fundamental synthesis parameters pH-value and precursor concentration in the
energy efficient microwave-assisted hydrothermal synthesis process of LiMnPO$_4$. Upon variation of these parameters,
we observe regimes were the product consists of well-dispersed particles shaped as hexagonal prisms which size is
adjustable from several $\mu$m down to the sub-$\mu$m range. In addition, we found a region of parameter space where
the particles exhibit a strong agglomerative behaviour, thus forming plate-like aggregates. The analysis of the XRD
patterns reveals preferred orientation of the particles and an anisotropic tendency to agglomeration, favouring
oriented aggregation in the ($b$,$c$)-plane. These results used to construct the agglomeration phase diagram which is
discussed with respect to the particles surface charging behaviour. In summary, appropriate choice of the pH-value and
the precursor concentration enables tailoring the size and morphology of LiMnPO$_4$ micro- and nanostructure as well as
controlling the dispersion or agglomeration.

\acknowledgement

The authors thank I. Glass for experimental support. Financial support by Bundesministerium f{\"u}r Bildung und Forschung
via the LIB2015 alliance (project 03SF0340/03SF0397) and by the DFG (KL 1824/2 and KL1824/5) is gratefully
acknowledged.

\begin{suppinfo}

Powder XRD patterns of all samples discussed and SEM images of the recovered precipitates are presented in the supporting information.

\end{suppinfo}


\begin{thebibliography}{10}
\bibitem{Vollath} Vollath, D. Nanomaterials: An Introduction to Synthesis, Properties and Applications. \emph{Wiley-VCH}, \textbf{2008}
\bibitem{Kelly} Kelly, K.L.; Coronado, E.; Zhao, L.L.; Schatz, G.C. The Optical Properties of Metal Nanoparticles: The Influence of Size, Shape, and Dielectric Environment. \textit{J. Phys. Chem. B} \textbf{2003}, \textit{107}, 668-677.
\bibitem{Penn} Penn, R. L. Kinetics of Oriented Aggregation. \emph{J. Phys. Chem. B} \textbf{2004}, \emph{108}, 12707-12712.

\bibitem{Guo2008} Guo, Y.-G.; Hu, J.-S.; Wan, L.-J. Nanostructured Materials for Electrochemical Energy Conversion and Storage Devices. \textit{Adv. Mater.} \textbf{2008}, \textit{20(15)}, 2878-2887.
\bibitem{Zakharova2012} Zakharova, G. S.; J{\"a}hne, C.; Popa, A. I.; T{\"a}schner, Ch.; Gemming, Th.; Leonhardt, A.; B{\"u}chner, B.; Klingeler, R. Anatase Nanotubes as an Electrode Material for Lithium-Ion Batteries. \textit{J. Phys. Chem. C} \textbf{2012}, \textit{116(15)}, 8714-8720.

\bibitem{Bruce} Bruce, P.G.; Scrosati, B.; Tarascon, J.M. Nanomaterials for Rechargeable Lithium Batteries. \textit{Angew. Chem. Int. Ed.} \textbf{2008}, \textit{47}, 2930-2946.
\bibitem{Armand2002} Armand, M.; Gauthier, M.; Magnan, J.-F.; Ravet, N. Method for synthesis of carbon-coated redox materials with controlled size. World Patent WO 02/27823 A1 \textbf{2001}.
\bibitem{Popa} Popa, A. I.; Vavilova, E.; T{\"a}schner, Ch.; Kataev, V.; B{\"u}chner, B.; Klingeler, R. Electrochemical Behavior and Magnetic Properties of Vanadium Oxide Nanotubes. \textit{J. Phys. Chem. C} \textbf{2011}, \textit{115(13)}, 5265-5270.
\bibitem{Padhi1997} Padhi, A. K.; Nanjundaswamy, K. S.; Goodenough, J. B. Phospho-olivines as Positive-Electrode Materials for Rechargeable Lithium Batteries. \textit{J. Electrochem. Soc.} \textbf{1997}, \textit{144(4)}, 1188-1194.
\bibitem{Yamada2001} Yamada, A.; Chung, S.-C. Crystal Chemistry of the Olivine-Type Li(Mn$_y$Fe$_{1-y}$)PO$_4$ and  (Mn$_y$Fe$_{1-y}$)PO$_4$ as Possible 4 V Cathode Materials for Lithium Batteries. \textit{J. Electrochem. Soc.} \textbf{2001}, \textit{148(8)}, A960-A967.
\bibitem{Amine2000} Amine, A.; Yasuda, H.; Yamachi, M. Olivine LiCoPO$_4$ as 4.8 V  Electrode Material for Lithium Batteries. \textit{Electrochem. Solid-State Lett.} \textbf{2000}, \textit{3(4)}, 178-179.
\bibitem{Okada2001} Okada, S.; Sawa, S.; Egashira, M.; Yamaki, J.; Tabuchi, M.; Kageyama, H.; Konishi, T.; Yoshino, A. Cathode properties of phospho-olivine LiMPO$_4$ for lithium secondary batteries. \textit{J. Power Sources} \textbf{2001}, \textit{97/98}, 430-432.
\bibitem{Morgan} Morgan, D.; Van der Ven, A.; Ceder, G. Li Conductivity in Li$_x$MPO$_4$ (M = Mn, Fe, Co, Ni) Olivine Materials. \textit{Electrochem. Solid-State Lett.} \textbf{2004}, \textit{7(2)}, A30-A32.

\bibitem{Kim2010} Kim, J.; Seo, D.-H.; Kim, S.-W.; Park, Y.-U.; Kang, K. Mn based olivine electrode material with high power and energy. \textit{Chem. Commun.} \textbf{2010}, \textit{46}, 1305-1307.
\bibitem{HpnW} Wang, D.; Buqa, H.; Crouzet, M.; Deghenghib, G.; Drezen, T.; Exnar, I.; Kwon, N.-H.; Miners, J.H.; Poletto, L.; Gratzel, M. High-performance, nano-structured LiMnPO$_4$ synthesized via a polyol method. \textit{J. Power Sources} \textbf{2009}, \textit{189(1)}, 624-628.
\bibitem{Drezen} Drezen, T.; Kwon, N-H.; Bowen, P.; Teerlinck, I.; Isono, M.; Exnar, I. Effect of particle size on LiMnPO$_4$ cathodes. \emph{Journal of Power Sources} \textbf{2007}, \emph{174} 949-953

\bibitem{GsaJ} Ji, H.; Yang, G.; Ni, H.; Roy, S.; Pinto, J.; Jiang, X. General synthesis and morphology control of LiMnPO$_4$ nanocrystals via microwave-hydrothermal route.  \textit{Electrochim. Acta} \textbf{2011}, \textit{56(9)}, 3093-3100.
\bibitem{Wizent} Wizent, N.; Behr, G.; Lipps, F.; Hellmann, I.; Klingeler, R.; Kataev, V.; L{\"o}ser, W.; Sato, N.; B{\"u}chner, B. Single-crystal growth of LiMnPO$_4$ by the floating-zone method. \textit{J. Cryst. Growth} \textbf{2009}, \textit{311(5)}, 1273-1277.
\bibitem{Li2006} Li, C.-C.; Lee, J.-T.; Peng, X.-W. Homogeneity and Cell Performance of Aqueous-Processed LiCoO$_2$ Cathodes by Using Dispersant of PAA - NH$_4$. \textit{J. Electrochem. Soc.} \textbf{2006},  \textit{153(5)}, A809-A815.
\bibitem{Toft} Toft-Petersen R.; Andersen N.H.; Li H.F.; Li J.Y.; Tian W.; Bud'ko S.L.; Jensen T.B.S.; Niedermayer C.; Laver M.; Zaharko O.;, Lynn J.W.; Vaknin D. Magnetic phase diagram of magnetoelectric LiMnPO4. \emph{Phys. Rev. B} \textbf{2012} 85,
224415.
\bibitem{Fiebig} van Aken, B.B.; Rivera, J. P.; Schmid, H.; Fiebig, M. Observation of ferrotoroidic domains. \emph{Nature} \textbf{2007} 449,
702-705.

\bibitem{Jaehne} J{\"a}hne, C.; Klingeler, R. Microwave-assisted hydrothermal synthesis of low-temperature LiCoO$_2$. \textit{Solid State Sci.} \textbf{2012}, \textit{14(7)}, 941-947.
\bibitem{fullprof} Rodriguez-Carvajal, J. Recent advances in magnetic structure determination by neutron powder diffraction. \textit{Physica B} \textbf{1993}, \textit{192(1-2)}, 55-69.
\bibitem{Carling1995} Carling, S.G.; Day, P.; Visser, D. Crystal and Magnetic Structures of Layer Transition Metal Phosphate Hydrates. \textit{Inorg. Chem.} \textbf{1995}, \textit{34}, 3917-3927.
\bibitem{IotG} Garcia-Moreno, O.; Alvarez-Vega, M.; Amador, U. Influence of the Structure on the Electrochemical Performance of Lithium Transition Metal Phosphates as Cathodic Materials in Rechargeable Lithium Batteries: A New High-Pressure Form of LiMPO$_4$ (M = Fe and Ni). \textit{Chem. Mater.} \textbf{2001}, \textit{13(5)}, 1570-1576.
\bibitem{DkvZ} Zemann, J. Die Kristallstruktur von Lithiumphosphat, Li$_3$PO$_4$. \textit{Acta Cryst.} \textbf{1960}, \textit{13}, 863-867.
\bibitem{CsoW} Waldrop, L. Crystal structure of triploidite. \textit{Naturwissenschaften} \textbf{1968}, \textit{55(6)}, 296-297.
\bibitem{Bramnik2007} Bramnik, N.N.; Ehrenberg, H. Precursor-based synthesis and electrochemical performance of LiMnPO$_4$. \textit{J. Alloys Compd.} \textbf{2008}, \textit{464}, 259-264.
\bibitem{Liu2012} Liu, J.; Hu, D.; Huang, T.; Yu, A. Synthesis of flower-like LiMnPO$_4$/C with precipitated NH$_4$MnPO$_4\cdot$H$_2$O as precursor. \textit{J. Alloys Compd.} \textbf{2012}, \textit{518}, 58-62.
\bibitem{Delacourt} Delacourt, C.; Laffont, L.; Bouchet, R.; Wurm, C.; Leriche, J.-B.; Morcrette, M.; Tarascon, J.-M.; Masquelier, C. Toward Understanding of Electrical Limitations (Electronic, Ionic) in LiMPO$_4$ (M = Fe, Mn) Electrode Materials. \textit{J. Electrochem. Soc.} \textbf{2005}, \textit{152(5)}, A913-A921.
\bibitem{Chen} Chen, G.; Richardson, T. J. Solid Solution Phases in the Olivine-Type LiMnPO$_4$/MnPO$_4$ System. \textit{J. Electrochem. Soc.} \textbf{2009}, \textit{156(9)}, A756-A762.
\bibitem{Rector} Rector, D. R.; Bunker, B. C. Effect of Colloidal Aggregation on the Sedimentation and Rheological Properties of Tank Waste. Report No. PNL-10761, Pacific Norwest Laboratories: \textbf{1995}, DOI: 10.2172/113874.
\bibitem{Li2007} Li, C.-C.; Lee, J.-T.; Tung, Y.-L.; Yang, C.-R. Effects of pH on the dispersion and cell performance of LiCoO$_2$ cathodes based on the aqueous process. \textit{J. Mater. Sci.} \textbf{2007}, \textit{42}, 5773-5777.
\bibitem{Mileni2007} Mileni, M.-P. Self-Assembly of Inorganic Nanocrystals: Fabrication and Collective Intrinsic Properties. \textit{Acc. Chem. Res.} \textbf{2007}, \textit{40}, 685-693.
\bibitem{Li2012} Li, J.; Armstrong, B. L.; Kiggans, J.; Daniel, C.; Wood, D. L. Optimization of LiFePO$_4$ Nanoparticle Suspensions with Polyethyleneimine for Aqueous Processing. \textit{Langmuir} \textbf{2012}, \textit{28}, 3783-3790.
\bibitem{Lee2010} Lee, J.-H.; Kim, H.-H.; Kim, G.-S.; Zang, D.-S.; Choi, Y.-M.; Kim, H.; Yi, D.K.; Sigmund, W.M.; Paik, U. Evaluation of Surface Acid and Base Properties of LiFePO$_4$ in Aqueous Medium with pH and Its Electrochemical Properties. \textit{J. Phys. Chem. C} \textbf{2010}, \textit{114}, 4466-4472.
\bibitem{CoiD} Dollase, W. A. Correction of intensities for preferred orientation in powder diffractometry: application of the March model. \textit{J. Appl. Cryst.} \textbf{1986}, \textit{19}, 267-272.
\bibitem{Xia2009} Xia, Y.; Xiong, Y.; Lim, B.; Skrabalak, S. E. Shape-Controlled Synthesis of Metal Nanocrystals: Simple Chemistry Meets Complex Physics? \textit{Angew. Chem. Int. Ed.} \textbf{2009}, \textit{48(1)}, 60-103.
\bibitem{Su2012} Su, J.; Wu, X.-L.; Yang, C.-P.; Lee, J.-S.; Kim, J.; Guo, Y.-G. Self-Assembled LiFePO$_4$/C Nano/Microspheres by Using Phytic Acid as Phosphorus Source. \textit{J. Phys. Chem. C} \textbf{2012}, \textit{116}, 5019-5024.

\end{thebibliography}
\end{document}